# A Categorization of Transparency-Enhancing Technologies

Christian Zimmermann
zimmermann@iig.uni-freiburg.de
Institute of Computer Science and Social Studies
University of Freiburg

**Abstract**

A variety of Transparency-Enhancing Technologies has been presented during the past years. However, investigation of frameworks for classification and assessment of Transparency-Enhancing Technologies has lacked behind. The lack of precise classification and categorization approaches poses an obstacle not only to systematic requirements analysis for Transparency-Enhancing Technologies but also to investigation and analysis of their capabilities and their suitability to contribute to privacy protection. This paper addresses this research gap. In particular, it presents a set of categorization parameters for describing the properties and functionality of a Transparency-Enhancing Technology on the one hand, and a categorization of Transparency-Enhancing Technologies on the other hand.

## 1. Transparency-Enhancing Technologies

The prevailing business model on the Internet relies on targeted advertising, which builds upon user profiling (Enders et al., 2008). The extensive data collection and analysis necessary for user profiling threaten users' privacy (Hildebrandt, 2009; Schermer, 2011) and have lead to an increased demand for privacy protection mechanisms. Classical privacy protection mechanisms aim primarily at confidentiality, i.e., data minimization and obfuscation (PISA Consortium, 2003). While these Privacy-Enhancing Technologies (PET) offer to users valuable mechanisms to protect themselves from unwanted data disclosure, they are not very well suited in the context of user profiling and personalization in Internet services (Müller et al., 2012; Novotny and Spiekermann, 2013; Solove, 2008; Weitzner et al., 2006). On the one hand, users do not primarily aim at data-minimization but are willing to disclose data in order to benefit from service usage and personalization (Acquisti and Gross, 2006; Fujitsu Res. Inst., 2010; Gross and Acquisti, 2005). On the other hand, PET can hardly prevent profilers from collecting a user's data from other sources than the user herself and can also hardly prevent inferences from already collected data (Accorsi et al., 2012; Calandrino et al., 2011). Currently, personal data is collected, analyzed and used by providers of Internet services based upon informed consent. However, the relation of users and providers of data-collecting services on the Internet is characterized by high information asymmetry (Schermer, 2011), i.e., users usually can not determine what data about them is actually collected by providers, let alone, which information about them is inferred from collected data and how that data is being used.
Transparency-Enhancing Technologies (TETs) aim at reducing this information asymmetry by providing users with information regarding providers' data collection, analysis and usage. Hansen defines TETs as "tools which can provide to the individual concerned clear visibility of aspects relevant to [its personal] data and the individual's privacy" (Hansen, 2008, p. 191). Hence, TETs do not aim at data-minimization but at providing users with insight into a data controller's intended and actual data handling behavior. Thus, TETs have the potential to constitute valuable complements to PET. A variety of very different TETs has been presented during the last years, e.g., the "DataTrack" (Fischer-Hübner et al., 2011), Google's "My Account"[1] or the "Mozilla Privacy Icons"[2]. However, only a few attempts at systematic categorization of TETs have been conducted. The resulting lack of a comprehensive classification framework for TETs impedes not only systematic requirements analysis for TETs but also analysis of their capabilities and their suitability to contribute to privacy protection. A common language to precisely describe a TET's properties, requirements and functionality is needed for the promotion of TET research and development. This paper addresses this research gap. In particular, it presents a set of categorization parameters for describing a TET's properties and functionality on the one hand, and a categorization of TETs based upon these parameters on the other hand.

The next section provides an overview on existing approaches towards categorization and classification of TETs. Section 3 presents the TET categorization "TetCat" and its underlying categorization parameters. The section further illustrates the development process of the TetCat and elaborates on both the underlying categorization parameters and the identified TET categories. Section 4 critically discusses the TetCat and compares it to the existing approaches introduced in Section 2 to illustrate this paper's contribution to the state of the art. Section 5 concludes the paper.

## 2. Existing Approaches

Only few work towards categorization of TETs has been conducted in this young research area. Four publications that provide categorization or classification approaches for TETs can be identified: (Hedbom, 2009), (Hildebrandt, 2009), (Janic et al., 2013) and (Zimmermann et al., 2014). In the following a brief overview over these existing approaches is provided. A more detailed illustration of the existing approaches is provided in the next section. Section 4 critically discusses the TetCat and compares it to the existing approaches.

In the context of "Ambient Intelligence" or the "Internet of Things", Hildebrandt distinguishes two types of TETs:

- **Type A:** "legal and technological instruments that provide (a right of) access to data processing, implying a transfer of knowledge from data controller to data subject, and / or"
- **Type B:** "legal and technological instruments that (provide a right to) counter-profile the smart environment to 'guess' how one's data match relevant group profiles that may affect one's risk and opportunities, implying that the observable and machine readable behaviour of one's environment provides enough information to anticipate the implications of one's behavior." (Hildebrandt, 2009)

Hildebrandt's categorization of TETs has the advantage of including not only technological instruments but also legal approaches towards transparency. However, this categorization does not provide further insight into technical functionality of a TET and properties such as a TET's trustworthiness. Hence, it is too coarse-grained to be suited for the contexts of development and assessment of technological TETs.

Hedbom's approach to categorizing TETs focuses more on the technological aspects of TETs (Hedbom, 2009). For classifying TETs, Hedbom proposes the classification parameters depicted in Table 1. While Hedbom's classification approach is more comprehensive and fine-grained than Hildebrandt's approach it suffers from several shortcomings. On the one hand, the classification parameter "Other Aspects" and its sub-parameters "Technologies Used" and "Security Requirements" are not clearly defined. Instead of allowing for precise classification of TETs they rather allow for a broad description of TETs. While this undoubtedly has its merits, it does not allow for easy comparison or categorization of TETs. Further, while Hedbom includes a trust-related classification parameter into its classification approach, he does not consider untrusted TETs. This seems rational, as untrusted TETs can hardly constitute valuable tools for supporting data-subjects in protecting their privacy. However, many tools, e.g. Google's My Account, exist that fall under Hansen's definition of TETs (Hansen, 2008), which is also used by Hedbom, but fit in none of Hedbom's manifestations of his "Trust Requirements" classification parameter. However, the parameter can be used to determine which trust requirements a TET would have to fulfill "to be considered as a transparency tool" (Hedbom, 2009, p.77). Hedbom's approach further does not distinguish between different sources of threats to privacy that a TET addresses. For example, My Account provides information on a data-controller's data collection and inferences, while the "Personal Information Dashboard" by Buchmann et al. also aims at transparency with regard to data disclosure to other users of a social network service (Buchmann et al., 2013).

---
[1] https://myaccount.google.com/intro
[2] https://wiki.mozilla.org/Privacy_Icons

Table 1 Hedbom's Classification Parameters (Hedbom, 2009)

| Classification Paramter | | Possible Manifestations |
|---|---|---|
| Possibilities of Control and Verification | | Promises |
| | | Read-Only |
| | | Interactive |
| Target Audiences | | Tools for Data Subjects |
| | | Tools for Auditors / Proxies |
| Scope | | Service Scope |
| | | Organizational Scope |
| | | Conglomerate Scope |
| Trust Requirements | | Trusted Server |
| | | Trusted Third Party |
| | | Trusted Client |
| | | No Trust Needed |
| Information Presented | | Required Information |
| | | Extended Information |
| | | Third Party Information |
| Other Aspect | Technologies Used | n.a. |
| | Security Requirements | Sensitivity of data |
| | | Concentration of data |
| | | Ease of access |

An overview over a selection of current TETs is presented by Janic et al., along with a high-level categorization of TETs (Janic et al., 2013). The presented TET categories are defined based on the transparency functionality they provide. Janic et al. identify the following five categories of transparency functionality:

- Transparency as insight in intended data collection, storage, processing and/or disclosure
- Transparency as insight in collected and/or stored data
- Transparency as insight in third party tracking (insight in user behaviour data disclosure)
- Transparency as insight in data collection, storage, processing and/or disclosure based on website's reputation

- Transparency as insight into (possibly) unwanted user's data disclosure (awareness promoting)

While this categorization provides some information on TETs, it does, e.g., not provide insight into the types of data regarding which a TET provides transparency. Further, it does not provide information regarding the trustworthiness of the information provided to data-subjects via a TET. Generally, Janic et al.'s categorization does provide information regarding what type of transparency a TET provides, but no details on how this transparency is provided with respect to technology.

Zimmermann et al. provide a specialized classification scheme for privacy dashboards, a specific class of TET (Zimmermann et al., 2014). Its classification parameters and their possible manifestations are depicted in Table 2.

Table 1 Zimmermann et al.'s classification parameters (Zimmermann et al., 2014)

| Classification Parameter | Manifestation | |
|---|---|---|
| Authentication Level | Anonymous | |
| | Pseudonymous | |
| | Fully Identified | |
| Data Types | Volunteered Data | |
| | Incidental Data | |
| | Observed Data | |
| | Derived Data | |
| Interactivity Level | Read-Only | |
| | Interactive | Collection |
| | | Modification |
| | | Deletion |
| | | Usage |
| Reach | One-to-One | |
| | One-to-Many | |
| Execution Environment | Client | |
| | Server | |
| | Hybrid | |
| | TTP | |
| Delivery Mode | Push | |
| | Pull | |
| | Push & Pull | |

The classification scheme allows for a relatively precise classification and description of privacy dashboards but it also suffers from several drawbacks. First, the approach is tailored specifically to privacy dashboards that provide information "… over data a data controller has accumulated about [data subjects]" (Zimmermann et al., 2014, p. 153). Hence, it is too specific for the purpose of categorization and classification of TETs in general. Second, similar to the above-presented approaches, it does not consider a TET's trustworthiness and different sources of privacy threats addressable by TETs.

As described above, the approaches by Hedbom, Hildebrandt, Zimmermann et al. and Janic et al. are not comprehensive and detailed enough to allow for detailed classification and assessment of TETs from a technical perspective. Hence, in the next section, a more comprehensive and detailed classification and categorization approach for TET is proposed.

# 3. Categorizing Transparency-Enhancing Technologies

In the following, categorization parameters for precisely describing a TET are provided. These categorization parameters serve two purposes: First, they provide a means to describe a TET's properties and functionalities and, second, they constitute a means to define distinct categories of TETs into which a specific TET can be classified. These categories are presented in Section 3.2.

## 3.1. Categorization Parameters

Figure 1 presents the categorization parameters building the foundation of the presented TET categorization TetCat. Further, Figure1 illustrates the development process of the TetCat's categorization parameters. As depicted, the parameters were constructed by synthesizing and extending the categorization approaches by Hedbom and Zimmermann et al. (Hedbom, 2009; Zimmermann et al., 2014).

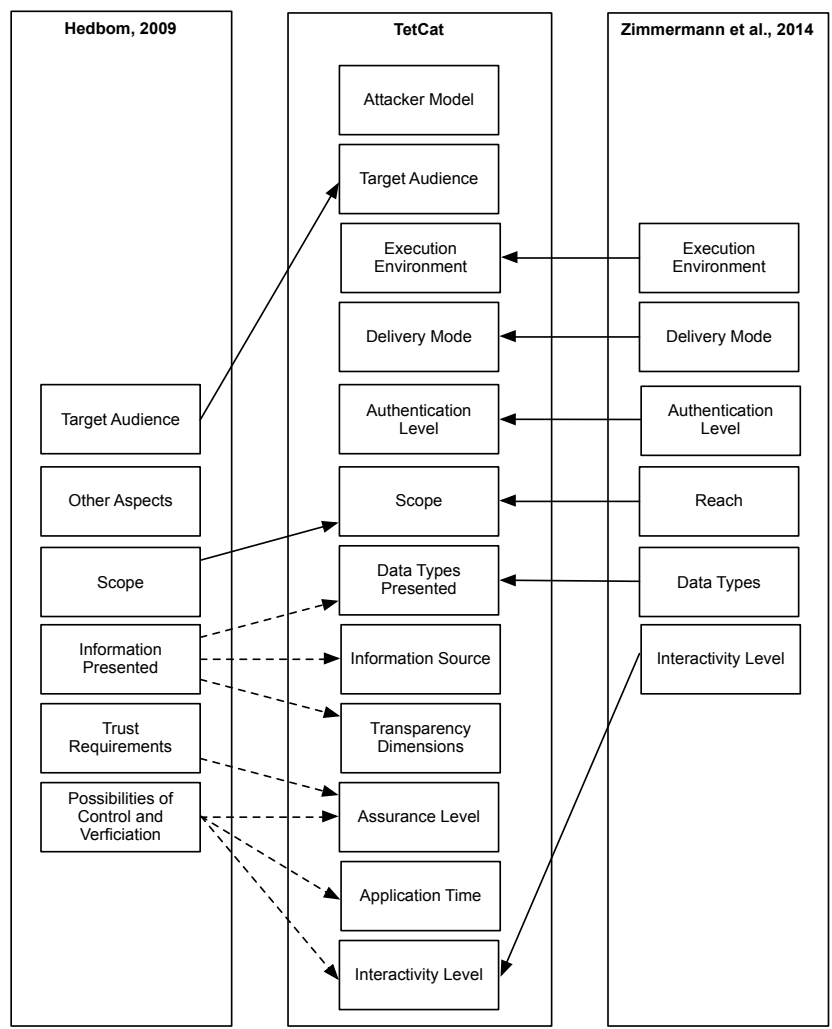

Figure 1 The TetCat's categorization parameters and their origins

In Figure 1, solid arrows indicate synthesis, adoption and/or extension of existing parameters and dashed arrows indicate implicit consideration. Hedbom's "Other Apects" have not been taken into account in the development of the TetCat, as they are not aimed at classification but constitute "rather a list of possible capabilities" respectively "list [of ] some aspects" (Hedbom, 2009, p. 74). In the following, the TetCat's categorization parameters are presented and the development process of the TetCat is illustrated. Figure 2 depicts the TetCat's categorization parameters and their possible manifestations.

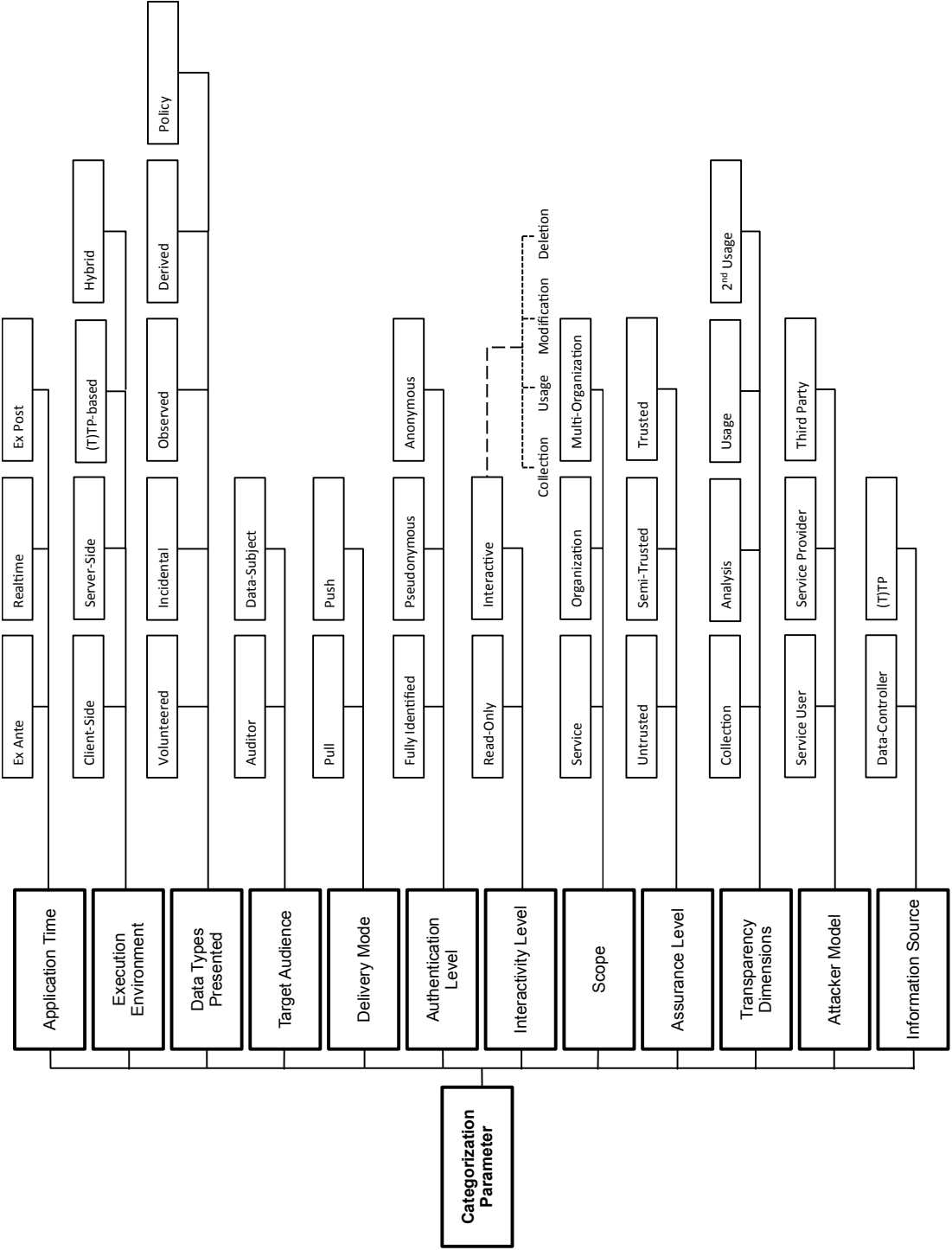

Figure 2 The TetCat's categorization parameters and their possible manifestation

### Application Time (AT)

The "Application Time" parameter describes the point of time at which a TET provides data-subjects with transparency, relative to the time of data collection resp. processing. This parameter implicitly draws from Hedbom's "Possibility of Control and Verification" (Hedbom, 2009).

TETs can be applied before, during or after data collection and processing. TETs that are applied ex ante data collection and processing provide information regarding a data-controller's reputation or her intended data collection and processing, e.g., by presenting to data-subjects easily understandable representations of the data-controller's privacy policy.

Hence, *Ex Ante* TETs are related to Janic et al.'s TETs that provide transparency "as insight in intended data collection, storage, processing and/or disclosure" and Hedbom's *Promises* TETs. Examples of TETs that aim at providing transparency ex ante data collection and processing are early versions of the DataTrack (Hansen, 2008) or the Mozilla Privacy Icons (see above).

TETs can also provide transparency during data collection and processing, e.g., by providing data-subjects with a real-time visualization of companies that follow them on the Internet via cookies. An example for a TET that is applied during data collection and processing is "LightBeam"[3].

While data collection and processing usually occur continuously, atomic steps in the process of data collection and processing can be identified, e.g., the collection of a specific data item at a specific point in time or the point of time at which a specific data item is derived from already collected data. Hence, TETs can provide insight into data collection and processing retrospectively, e.g., by providing data-subjects with insight into data about them stored by a data-controller. Examples of TETs that provide transparency ex post data collection and processing are current versions of the DataTrack (Fischer-Hübner et al., 2011), My Account or Acxiom's "AboutTheData" portal[4].

### Target Audience (TA)

Following Hedbom, the TetCat distinguishes between TETs that directly address data-subjects and TETs that address external auditors (Hedbom, 2009).

### Interactivity Level (IL)

The "Interactivity Level" parameter describes, from a technological perspective, the level of control that a data-subject can exercise via a TET. While *Read-Only* TETs provide data-subjects solely with insight into a data-controller's data collection and processing, *Interactive* TETs allow data-subjects to exercise control over collection and processing through negotiation of privacy policies, through exercising control over usage or through allowing deletion or modification of information from existing profiles (Hedbom, 2009; Zimmermann et al., 2014). An example of a *Read-Only* TET is "Privacy Bird"[5], while My Account is an example of an *Interactive* TET.

The IL parameter is adopted from (Zimmermann et al., 2014). Hedbom's "Possibilities of Control and Verification" also resembles this parameter. However, "Interactivity Level" does not include Hedbom's *Promises* manifestation. *Promises* TETs provide information on data-controllers' intended or purported behavior but "give no on line [sic!] or automatic way for the user/proxy to verify these claims" (Hedbom, 2009, p. 70). Hence, *Promises* TETs do not provide data-subjects with technological means to exercise control. Thus, in this paper, *Promises* TETs are considered *Read-Only* TETs and the information on whether a TET provides insight into actual or intended behavior is considered in the parameters "Assurance Level" and "Application Time".

---
[3] https://www.mozilla.org/de/lightbeam/

[4] https://www.aboutthedata.com

[5] http://www.privacybird.org

**Delivery Mode (DM)**

Following Zimmermann et al.'s approach, the "Delivery Mode" parameter describes how a TET notifies data-subjects of aspects relevant to their privacy. Transparency-Enhancing Technologies can either actively notify data-subjects (*push*) of events relevant to their privacy or require data-subjects to actively inform themselves (*pull*) (Zimmermann et al., 2014). For example, Ghostery[6] actively warns data-subjects of trackers (*push*) while current privacy dashboards, e.g., the AboutTheData portal, require data-subjects to actively visit the portal website to gain insight (*pull*).

**Authentication Level (AL)**

As the DM parameter, the parameter "Authentication Level" is adopted directly from (Zimmermann et al., 2014). Transparency-Enhancing Technologies can be used anonymously, pseudonymously or fully identified (Zimmermann et al., 2014). For example, tools such as Ghostery that solely visualize trackers that are sent by a website do not require data-subjects to authenticate themselves. Privacy dashboards provided by data-controllers or (Trusted) Third Parties ((T)TPs), however, require data-subjects to authenticate themselves in order to match data-subjects to their profiles (Zimmermann et al., 2014). While, e.g., My Account can be used pseudonymously, the AboutTheData portal requires data-subjects to fully identify themselves through their social security number.

**Execution Environment (EE)**

As the previous two parameters, the "Execution Environment" parameter is adopted straight forward from (Zimmermann et al., 2014).
This categorization parameter describes on which party's systems a TET is operated (Zimmermann et al., 2014). *Client-side* TETs are operated by data-subjects on their own systems, e.g., as a browser plug-in such as LightBeam. *Server-side* TET are operated by data-controllers which provide them to data-subjects, e.g., via a website such as Google's My Account.
TETs that are *(T)TP-based* are provided by (trusted) third parties to provide data-subjects with transparency with regard to specific data-controllers' data handling practices. If one considers the website's provider trustworthy, the "Me & My Shadow" website[7] can be considered a TTP-based TET. It provides data-subjects with information regarding which data they disclose when using specific web services, operating systems or other applications. Other examples of *(T)TP-based* TETs are government-supported privacy seals such as the "European Privacy Seal"[8].
*Hybrid* TETs combine traits of the aforementioned tools, i.e., they comprise different modules that are operated by different parties. Current versions of the DataTrack constitute hybrid TETs, as the underlying middleware is executed on the systems of data-subjects as well as on those of data-controllers (Fischer-Hübner et al., 2011).

---

[6] https://addons.mozilla.org/de/firefox/addon/ghostery
[7] https://myshadow.org
[8] https://www.european-privacy-seal.eu

### Scope (SP)

The "Scope" parameter synthesizes Zimmermann et al.'s "Reach" parameter and Hedbom's "Scope" parameter (Hedbom, 2009; Zimmermann et al., 2014). It describes the range of services respectively data-controllers that a TET considers in providing transparency. Transparency-Enhancing Technologies can aim at providing transparency with regard to a service, to an organization or to several organizations. A TET that exhibits a S*ervice* scope is, e.g., the transparency functionality of the Amazon Book Recommendation System, which provides Amazon's customers with information regarding the data Amazon has stored about them and how that data is being used for recommendations. A TET that provides transparency with respect to an organization is, e.g., the AboutTheData portal. An example for a TET with Scope *Multi-Organization* is the Personal Information Dashboard (Buchmann et al., 2013) that aims at providing data-subjects with transparency with regard to several Social Network Services (SNS).

### Data Types Presented (DT)

The "Data Types Presented" parameter is an enhancement of the "Data Types" parameter by Zimmermann et al. (Zimmermann et al., 2014). In contrast to the implicitly related "Information Presented" parameter from (Hedbom, 2009), it takes a technological perspective to describe regarding which types of data a TET provides insight.

The parameter's manifestations are based on (Schneier, 2010) and (World Economic Forum, 2012). In this work, as in (Zimmermann et al., 2014), *Volunteered Data* is defined as "data a user actively and knowingly discloses", *Observed Data* as "data a user passively discloses, i.e., data that results from the interaction of a user with a provider", *Incidental Data* as "data about a user that is disclosed not by the user herself but by others" and *Derived Data* as "data about users that is inferred as result of data analysis" (Zimmermann et al., 2014, p. 153). In contrast to Zimmermann et al.'s "Data Types" parameter, "Data Types Presented" further features the manifestation *Policy* to take into account TETs that do not provide insight into data already stored by a data-controller but also TETs that provide insight into a data-controller's privacy policy.

### Information Source (IS) and Transparency Dimension (TD)

Both the "Information Source" and the "Transparency Dimensions" parameters are implicitly connected to Hedbom's "Information Presented" parameter (Hedbom, 2009). The "Information Presented" parameter distinguishes between *Required information*, *Extended information* and *Third party information* (Hedbom, 2009, p. 73). The former two manifestations of Hedbom's parameter consider the information that a TET presents from a legal perspective. While *Required information* is information that a "service provider has to provide according to the Law [sic!]" (Hedbom, 2009, p.73), *Extended information* is information "that is not legally required" (Hedbom, 2009, p.73). In contrast to that perspective, the latter manifestation (*Third party information*) refers to the source of information from a technological perspective. It is used by Hedbom to describe TETs that "[gather] and [present] information given or harvested from other sources than the service provider" (Hedbom, 2009, p.73).

The TetCat avoids this mixing of different perspectives within one categorization parameter by building upon three categorization parameters to describe information source on the one hand and information extent on the other hand. While the parameter "Data Types Presented" (see above) describes information extent from a technological perspective, the parameter "Transparency Dimension" describes information extent with regard to the different stages of the data utilization cycle. First, TET can provide data-subjects with insight into data collection, i.e., insight into which data is collected by whom (*Collection*). Second, they can also provide data-subjects with insight into the analysis of the data collected by data-controllers, e.g., by providing data-subjects with information on inference antecedents (*Analysis*). Third, TETs can provide data-subjects with insight into how and for what purposes data about them is being used by a data-controller (*Usage*). Finally, TETs can provide data-subjects with insight into second usage of data, e.g., information on how third parties use data about them and the purposes and conditions of the data transfer from the original data-controller to these parties (*2$^{nd}$ Usage*).

The information provided by a TET can stem from different source. The "Information Source" parameter considers two potential sources of information that a TET can draw from to provide information. Data-Controllers themselves can provide data-subjects with information relating to a data-subject's privacy and her personal data via a TET (*Data-Controller*). A TET can, however, also provide information by drawing from other, trusted or untrusted, sources (*(T)TP*), such as government audit of privacy practices, privacy seals or (crowd-sourced) reputation services (Hedbom, 2009; Janic et al., 2013). The "Information Source" parameter also takes into account Janic et al.'s insight that TET can provide information stemming from sources other than the data-controller herself. They partly take this into account in their category of TETs that provide "transparency as insight in data collection, storage, processing and/or disclosure based on website's reputation" (Janic et al., 2013, p. 22).

### Assurance Level (AS)

This categorization parameter implicitly is related to Hedbom's "Possibilities of Control and Verification" and "Trust Requirements" parameters (Hedbom, 2009) (see above). It is used to describe the extent to which data-subjects can determine the completeness and correctness of the information that a TET provides them with. The correctness and completeness of the information provided by *Untrusted* TETs can not be determined by the data-subjects and no auditing entity that acts on the data-subject's behalf exists. *Trusted* TETs provide information, whose correctness and completeness are guaranteed by technical means. The correctness and completeness of the information provided by *Semi-trusted* TETs can not be guaranteed by technical means but a data-subject or an auditor can manually determine whether the information is correct and complete.

### Attacker Model (AM)

In providing transparency, TETs can focus on different sources of threats to data-subjects' privacy. TETs can focus on transparency within a specific service with respect to other users of the service, e.g., by providing insight into which other users of a SNS can see a data-subject's posts (*Service User*). Further, TETs can also focus on specific service providers a data-subject interacts with and provide the data-subject with, e.g., insight into stored profile data and its usage (*Service Provider*). Finally, TETs can focus on insight into third parties' data handling, e.g., that of parties that a data-subject does not interact with directly or knowingly, e.g., a service provider's advertising partners or companies crawling publicly accessible data on the Internet (*Third Party*).

## 3.2. Categories of Transparency-Enhancing Technologies

The above described categorization parameters can be utilized to define TET categories that differ with regard to the manifestations of specific categorization parameters. The here presented categorization TetCat depicted in Figure 3 comprises six categories of TETs.

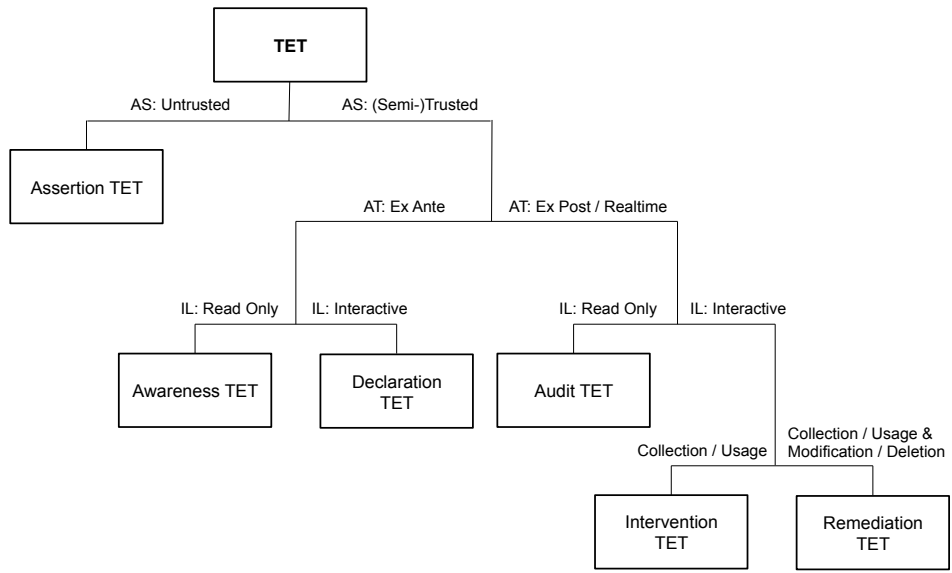

Figure 3 TetCat: Categorization of Transparency-Enhancing Technologies

For category definition the categorization parameters Assurance Level, Application Time and Interactivity Level are used as key determinants for distinction. The parameter Assurance Level is used as the primary determinant for defining TET categories, while Application Time and Interactivity Level constitute secondary, respectively tertiary, determinants. Assurance Level is used as primary determinant because a TET's suitability to support privacy protection heavily depends on its trustworthiness. Application Time has been chosen as secondary determinant to be able to clearly distinguish TETs that aim at enabling data-subjects to make informed decisions regarding whether to use or not to use a specific service and TETs that aim at enabling data-subject to gain insight into, and possibly exercise control over, already disclosed data. Finally, Interactivity Level has been chosen as tertiary determinant to be able to distinguish between TETs that only provide insight and TETs that empower data-subjects to exercise control over data relating to them. In the following, the categories are presented. A critical analysis of the TetCat is provided in Section 4.

## Assertion TETs

The TET category "Assertion TET" is defined by exhibiting Assurance Level *Untrusted*, i.e., Assertion TETs provide no means to determine the correctness and completeness of the information they provide. The TetCat does not further distinguish between *Untrusted* TETs based on other parameters. It refrains from doing so, as an *Untrusted* TET's unsuitability to provide trustworthy transparency remains unaffected by its manifestations of other parameters. Assertion TETs can only provide data-subjects with information regarding a data-controller's purported data handling practices.

## Awareness TETs and Declaration TETs

The categories "Awareness TET" and "Declaration TET" exhibit Assurance Level *Semi-Trusted* or *Trusted* and Application Time *Ex Ante*. However, while Awareness TETs exhibit Interactivity Level *Read-Only*, Declaration TETs are interactive.

As TETs of both categories are applied ex ante data collection and processing, they do not aim at providing transparency with regard to already existing profiles but with respect to data-controllers' intended data handling practices or their privacy policies or reputation.

The Assurance Level of Awareness TETs refers to data-subjects' (respectively auditors) capability to determine the correctness and completeness of the information provided about the privacy policy or the data-controllers' compliance with specific privacy requirements such as defined by entities that issue privacy seals. The same holds for Declaration TETs. However, they also provide data-subjects with functionality for policy negotiation.

### Audit TETs, Intervention TETs and Remediation TETs

Exhibiting Application Time *Ex Post*, the categories "Audit TET", "Intervention TET" and "Remediation TET" comprise TETs with Assurance Level *(Semi-)Trusted* that are able to provide data-subjects with insight into data actually stored by data controllers and into how the data is used by them. While Audit TETs exhibit Interactivity Level *Read-Only* and, hence, provide data-subjects only with insight into stored data and, possibly, its usage, Intervention TETs and Remediation TETs also provide data-subjects with functionality to exercise control over data stored on the data-controller side. Intervention TETs exhibit Interactivity Level *Interactive* and provide functionality to exercise control over further collection and/or usage of data. Remediation TETs also provide functionality to exercise control over further collection and/or usage of data but also comprise functionality for modification and/or deletion of data relating to a data-subject already stored by a data-controller.

## 4. Discussion

In order to discuss the TetCat and critically compare it to existing classification and categorization approaches, in the following it is applied to TETs that have already been classified in (Hedbom, 2009; Janic et al., 2013; Zimmermann et al., 2014). In particular, the TETs Privacy Bird, MyAccount, Lightbeam (formerly named "Collusion") and the Amazon Book Recommendation System are classified in the following.

Tables 4a and 4b depict the analysis results. In the tables, green fields indicate that the respective classification is taken directly from the respective authors while white fields indicate that the respective classification was performed by the author of this paper based on the respective categorization approaches. The My Account constitutes a special case. The My Account TET is an amalgamation of the TETs "Google Privacy Dashboard" and "Google APM", which were analyzed in (Janic et al., 2013) and (Zimmermann et al., 2014), respectively. While, the original classification by Janic et al. remains unaffected by the amalgamation, the respective fields in Table 4a are still colored white to indicate the reapplication of their approach. Due to the amalgamation, Zimmermann et al.'s original classification of the Google APM can not be compared to a classification of the My Account. Thus, their approach has also been reapplied to the MyAccount. As can be seen in Tables 4a and 4b, the categorization approach by Janic et al. allows for far less fine-grained description and categorization of TETs than the other approaches. Hence, their approach is not further considered in this discussion.

The approach by Zimmermann et al. focuses specifically on privacy dashboards and, hence, is not applicable to TET that do not provide information "… over data a data controller has accumulated about [data subjects]" (Zimmermann et al., 2014, p. 153). Consequently, the TetCat is better suited to allow for classification of TETs in general.

| TET | Janic et al. | Hedbom | | Zimmermann et al. | | TetCat | | |
|---|---|---|---|---|---|---|---|---|
| | | Categorization Parameter | Manifestation | Categorization Parameter | Manifestation | Categorization Parameter | Manifestation | Category |
| Privacy Bird | "Transparency as insight in intended data collection, storage, processing and/or disclosure" | Possibilities of Control and Verification | Promises | Not Applicable | | Application Time | Ex Ante | Awareness TET |
| | | | | | | Execution Environment | Hybrid | |
| | | | | | | Data Types Presented | Policy | |
| | | Target Audience | Data-Subjects | | | Target Audience | Data-Subject | |
| | | | | | | Delivery Mode | Push | |
| | | | | | | Authentication Level | Anonymous | |
| | | | | | | Interactivity Level | Read-Only | |
| | | Scope | Service | | | Scope | Service | |
| | | Trust Requirements | Would need a Trusted Server to be a transparency tool | | | Assurance Level | Semi-Trusted | |
| | | Information Presented | Required Information | | | Transparency Dimensions | Collection, Analysis, Usage, partially 2nd Usage | |
| | | | | | | Attacker Model | Service-Provider | |
| | | | | | | Information Source | Data-Collector | |
| Google Dashboard / MyAccount | "Transparency as insight in collected and/or stored data" | Possibilities of Control and Verification | Interactive | Authentication Level | Pseudonymous, Fully Identified | Application Time | Realtime, Ex Post | Assertion TET |
| | | | | | | Execution Environment | Server-Side | |
| | | | | Data Types | Volunteered, Derived, Observed | Data Types Presented | Volunteered, Derived, Observed | |
| | | Target Audience | Data-Subjects | | | Target Audience | Data-Subject | |
| | | | | Interactivity Level | Interactive (Collection, Modification, Deletion) | Delivery Mode | Pull | |
| | | | | | | Authentication Level | Pseudonymous, Fully Identified | |
| | | | | | | Interactivity Level | Interactive (Modification, Deletion, Collection) | |
| | | Scope | Organizational | Reach | One-to-One | Scope | Organization | |
| | | Trust Requirements | Would require a Trusted Server | Execution Environment | Server-Side | Assurance Level | Untrusted | |
| | | Information Presented | Required, Extended | | | Transparency Dimensions | Collection, Analysis, partially Usage | |
| | | | | Delivery Mode | Pull | Attacker Model | Service-Provider | |
| | | | | | | Information Source | Data-Collector | |

**Figure 4a Comparison of the TetCat with existing approaches**

| TET | Janic et al. | Hedbom | | Zimmermann et al. | | TetCat | |
|---|---|---|---|---|---|---|---|
| | | Categorization Parameter | Manifestation | Categorization Parameter | Manifestation | Categorization Parameter | Manifestation |
| Lightbeam Collusion | "Transparency as insight in third party tracking (insight in user behaviour data disclosure)" | Possibilities of Control and Verification | Partially Interactive | Not Applicable | | Application Time | Realtime, Ex Post |
| | | | | | | Execution Environment | Client-Side |
| | | | | | | Data Types Presented | Observed |
| | | Target Audience | Data-Subjects | | | Target Audience | Data-Subjects |
| | | | | | | Delivery Mode | Pull |
| | | | | | | Authentication Level | Anonymous |
| | | | | | | Interactivity Level | Interactive (Collection) |
| | | Scope | Conglomerate | | | Scope | Multi-Organization |
| | | | | | | Assurance Level | Semi-Trusted |
| | | Trust Requirements | Trusted Third Party Needed | | | Transparency Dimensions | Collection, 2nd Usage |
| | | | | | | Attacker Model | Service-Provider, Third Party |
| | | Information Presented | Required Information, Third Party Information | | | Information Source | 3rd Party |
| Amazon Book Recomen- dation System | "Transparency as insight in collected and/or stored data" and partially "Transparency as insight into (possibly) unwanted user's data disclosure (awareness promoting)" | | | | | Application Time | Realtime, Ex Post |
| | | Possibilities of Control and Verification | Partially Interactive | Authentication Level | Fully Identified | Execution Environment | Server-Side |
| | | | | Data Types | Observed, Derived, Volunteered | Data Types Presented | Observed, Derived, Volunteered |
| | | Target Audience | Data-Subjects | | | Target Audience | Data-Subject |
| | | | | Interactivity Level | Interactive (Usage, Deletion, Modification) | Delivery Mode | Pull |
| | | | | | | Authentication Level | Fully Identified |
| | | | | | | Interactivity Level | Interactive (Deletion, Modification, Usage) |
| | | Scope | Service | Reach | One-To-One | Scope | Service |
| | | Trust Requirements | Would require Trusted Server | Execution Environment | Server-Side | Assurance Level | Untrusted (Derived), Semi-Trusted (Observed) |
| | | | | | | Transparency Dimensions | Collection, Analysis, partially Usage |
| | | | | | | Attacker Model | Service-Provider |
| | | Information Presented | Required or Extended | Delivery Mode | Pull | Information Source | Data-Collector |

(Assertion TET / Remediation TET category applies; Intervention TET category also shown.)

**Figure 4b Comparison of the TetCat with existing approaches**

The categorization approach by Hedbom does not provide categories of TETs but only classification parameters, which are suited to describe TETs to a limited extent. However, besides providing distinct categories of TETs, the TetCat allows for a more fine-grained and precise description of TETs because of its more fine-grained and detailed categorization parameters. For example, as can be seen in the classifications of the Amazon Recommendation System or the My Account TET, the TetCat allows for a more precise description of data-subjects' capability to interact with the system. As can be seen in the classification of the Privacy Bird, the TetCat is also better suited to describe the technological infrastructure of a TET (Execution Environment) and the provenance of the data presented by a TET (Information Source).

While the TetCat provides a more fine-grained categorization and allows for a more precise classification than current approaches towards classifying and categorizing TETs, it still exhibits some limitations. Obviously, the categorization depicted in Figure 3 is not completely final. More categories can be defined to allow for a more detailed categorization. For example, a TET can provide transparency before data collection and processing while at the same time also providing insight after collection and analysis. Further, a TET can also exhibit Assurance Level *Untrusted* with respect to some of the information it provides while at the same time providing other information whose correctness and completeness can be determined by a data-subject or auditor or is guaranteed by technical means. This is the case with the Amazon Book Recommendation System, which, as depicted in Table 4b, could be classified as both an Assertion TET and a Remediation TET.

However, this paper aims at providing an easily understandable and usable classification and categorization approach to lay the ground for a common terminology to facilitate requirements analysis, development and assessment of TETs. Hence, in order to preserve readability and usability, while at the same time providing a more comprehensive and detailed classification and categorization approach than currently available, no further categories are included into the TetCat.

Another limitation refers to the categorization parameter Interactivity Level. It refers to a TET's capability to enable data-subjects to exercise control over data stored on the side of a data-controller or to negotiate policies. However, as is the case with the Personal Information Dashboard (Buchmann et al., 2013), TET can also provide a limited extent of interactivity without enabling data-subjects to exercise control over a data-controller's data storage and handling. For example, the Personal Information Dashboard provides data-subjects with functionality to delete other users' posts from her Facebook wall. However, this does not necessarily delete the posts' content and the resulting inferences from Facebook's internal storage. This type of interactivity is not taken into account within the presented approach.

## 5. Conclusion

This paper aimed at facilitating requirements analysis for and development and assessment of Transparency-Enhancing Technologies. It presented a classification approach for Transparency-Enhancing Technologies and a categorization of these instruments in order to provide the basis for a common terminology. The presented approach builds upon previous work by Hedbom and on joint work by the author of this paper and colleagues (Hedbom, 2009; Zimmermann et al., 2014).

The applicability of the presented approach and its contribution to the state of the art have been demonstrated in a critical comparison with existing approaches.

Although it exhibits some limitations, the presented classification approach and categorization provide a means to categorize and describe Transparency-Enhancing Technologies in a more comprehensive and detailed manner than previous approaches.